\documentclass[aip,reprint,amssymb,twocolumn]{revtex4}  

\usepackage{dcolumn}% Align table columns on decimal point
\usepackage{bm}% bold math
\usepackage[utf8]{inputenc}
\usepackage[T1]{fontenc}
\usepackage{mathptmx}
\usepackage{etoolbox}

\usepackage{amsmath}% 
\usepackage{graphicx}%
\usepackage{color}%
\usepackage{lineno,hyperref}
\makeatletter
\def\@email#1#2{%
 \endgroup
 \patchcmd{\titleblock@produce}
  {\frontmatter@RRAPformat}
  {\frontmatter@RRAPformat{\produce@RRAP{*#1\href{mailto:#2}{#2}}}\frontmatter@RRAPformat}
  {}{}
}%
\makeatother

\begin{document}
%\begin{frontmatter}
\title{Revealing the role of $\Sigma$3\{112\} Si grain boundary local structures in impurity segregation}
\author{Rita Maji}
\address{Dipartimento di Scienze e Metodi dell'Ingegneria, Universit{\`a} di Modena e Reggio Emilia,Via Amendola 2 Padiglione Tamburini , I-42122 Reggio Emilia, Italy}
\author{Eleonora Luppi}
\address{Laboratoire de Chimie Th\'eorique, Sorbonne Universit\'e and CNRS  F-75005 Paris, France}
\author{Elena Degoli}\thanks{Corresponding author: elena.degoli@unimore.it }
\address{Dipartimento di Scienze e Metodi dell'Ingegneria, Universit{\`a} di Modena e Reggio Emilia and Centro Interdipartimentale En$\&$Tech, Via Amendola 2 Padiglione Morselli, I-42122 Reggio Emilia, Italy, \\ 
Centro S3, Istituto Nanoscienze-Consiglio Nazionale delle Ricerche (CNR-NANO),Via Campi 213/A, 41125 Modena, Italy \\
Centro Interdipartimentale di Ricerca e per i Servizi nel settore della produzione, stoccaggio ed utilizzo dell'Idrogeno H$2$–MO.RE., Via Universit{\`a} 4, 41121 Modena, Italy} 
\date{\today}

%\pacs{?}

\begin{abstract}
The interfacial structure of a silicon grain boundary (Si-GB) plays a decisive role on its chemical functionalization and has implications in diverse physical-chemical properties of the material. Therefore, GB interface is particularly relevant when the material is employed in high performance technological applications.
Here, we studied from first principles the role of GB interface by providing an atomistic understanding of two different $\Sigma$3\{112\} Si-GB models. These models are (1$\times$1) and (1$\times$2) $\Sigma$3\{112\} Si-GBs which lead to different structural reconstruction. Starting from these two models, we have shown that geometry optimization has an important role on the structural reconstruction of the GB interface and therefore on its properties. For this reason, we discussed different methodologies to define an optimal relaxation protocol. The influence of the local structures in (1$\times$1) and (1$\times$2) models have also been investigated in the presence of vacancies where different light impurities of different valency (C, N, H, O) can segregate. We studied how local structures  in (1$\times$1) and (1$\times$2) models are modified by the presence of vacancies and impurities. These structural modifications have been correlated with the changes of the energetics and electronic properties of the GBs. The behaviour of (1$\times$1) and (1$\times$2) models demonstrated to be significantly different. The interaction with vacancies and the segregation of C, N, H and O are significantly different depending on the type of local structures present in $\Sigma$3\{112\} Si-GB.

\end{abstract}

%\begin{keyword}
%end{keyword}

%\end{frontmatter}

\maketitle

\section{Introduction}\label{sec:introduction}
%basics on poly-Si & detrimental charaters
Polycrystalline silicon (poly-Si) is very attractive for photovoltaic applications. \cite{pvSi2015,aplSisolar2020}. However, its efficiency can be strongly modified because poly-Si is characterized by a high density of grain boundaries (GBs) \cite{Sisolarcell2009review,Sisolarcellthinfilm2004,Sisolarcellbook2019}, that can affect device performances  \cite{GBeffect1997,GBeffect1990,GBeffect2012} as 
lattice distortions is induced. This makes advantageous charge carriers recombination which induces a significant reduction of carrier lifetime. \cite{lifetime2005,PhysRevLett_115_235502,JAP_H2009,YuAndreyJAP2015} Furthermore, GBs easily form vacancies with deep defect electronic states. \cite{JAP_H2009,sisolarcell2020} Moreover, GBs facilitates the segregation of several impurities which modifies the electrical properties of poly-Si. \cite{SinnoJAP2015, MAJI2021116477,JCP_Rmaji2021,Majipssp21,ZHAO2017599,PhysRevB_91_035309,ohnoapl2013,KasinnoJAP13,YuAndreyJAP2015,OhnoAPL2017,OhnoAPL15,ShiJAP2010,PhysRevLett_121_015702} Therefore, recombination and segregation activity can have a substantial detrimental impact on the conversion efficiency of the solar cells. 
\cite{YuAndreyJAP2015,Peaker2012}
Another potential application of poly-Si, featuring as a channel is on 3D NAND devices, \cite{3DNAND_2018, 3DNAND_2022}
however, the conduction is significantly degraded due to the scattering at GBs and interface defects.

In order to optimize grain boundary effects on the solar cell performances, atomic hydrogen passivation is frequently considered: H is actually inherent during growth process.\cite{MARTINUZZI2003343,H_passivate2004,JAP_H2009} Moreover, light elements like C, N, O that exist as contaminants along GBs\cite{JCP_Rmaji2021}, can become efficient gettering center for other native defects.

% serach of GB structure : literature 
At present, some of the computational studies focuses to establish a generalised behaviour of GBs \cite{Lazebnykh2015,sisolarcell2020}, however the accurate correlation between the electrical and structural properties of a GB cannot be done considering
only the misorientation of two grains and the $\Sigma$ value
alone. A detailed analysis of the GB plane, local deviations in
the orientation as well as the presence of extrinsic defects at the atomic scale, is compulsory. Now, concerning the structural properties a reliable modeling of GBs is needed, as explored in more detail in Ref.\cite{Nature2021} considering the diversity of GBs. So far, in the literature, the density functional theory (DFT) has been considered the most reliable approach to optimize this type of supercells in addition to having a more reasonable computational cost, however one have to be very careful when considering asymmetric or defected cells where the DFT can incur in non realistic structural minima.
Actually, despite extensive computational and experimental investigations, presence of under- or over-coordination at interfaces in some GBs makes difficult to identify the most correct and realistic optimized structure for such complex systems: this is therefore still a source of debate.\cite{Nature2021,JMR2021} 

% our consideration & motivation
This work aims to a better understanding of the interaction between GB local structures and vacancies defects as well as impurity defects such as C, N, H and O. We have chosen for this investigation symmetric $\Sigma$3\{112\} GB which can be both under and over coordinated and, consequently, which can be described by two different possible optimized geometries. This systematic investigation of $\Sigma$3\{112\} GB was done in comparison with $\Sigma$3\{111\} GB that is instead perfectly four fold coordinated. 

The method to obtain the correct reconstructed GB geometry is described in section \ref{method} and the effect of vacancies on both the structural and electronic properties of the GBs are analysed in section \ref{sec:vacancy}.
In section \ref{sec:impurity} we described the segregation of impurities and their electronic structures in the two GB models. The discussion about the correlation between GBs local structures and vacancies/impurities, together with the comparison with $\Sigma$3\{111\} GB, is in section \ref{sec:discussion}). 

\section{Methodology}
\label{method}

The calculations were performed using density functional theory (DFT) as implemented in the plane-wave based Vienna Ab initio Simulation Package (VASP).\cite{Hafner, Kresse} We employed the generalised gradient approximation PBE for the exchange-correlation functional \cite{PhysRevLett.77.3865} and projector augmented-wave (PAW) pseudopotentials with a cutoff of 400 eV. K-points sampling within the Monkhorst Pack scheme \cite{Monkhorst} was used for integration of Brillouin-zone together with the linear tetrahedron method including Bl\"ochl corrections. \cite{PhysRevB.49.16223} In particular, we used a k-mesh of 5$\times$5$\times$3 to calculate energetics of the structures and a k-mesh of 7$\times$7$\times$5 to calculate their density of states (DOS). For the structural optimisation, we used a force threshold value of 10$^{-2}$ eV/\AA{} per atom. The analysis of the geometrical structures and post-processing analyses were carried out using the utilities of VESTA \cite{Momma_vesta}.

%============================================================================
\begin{figure*}[t]
\centering
\includegraphics[scale=0.22]{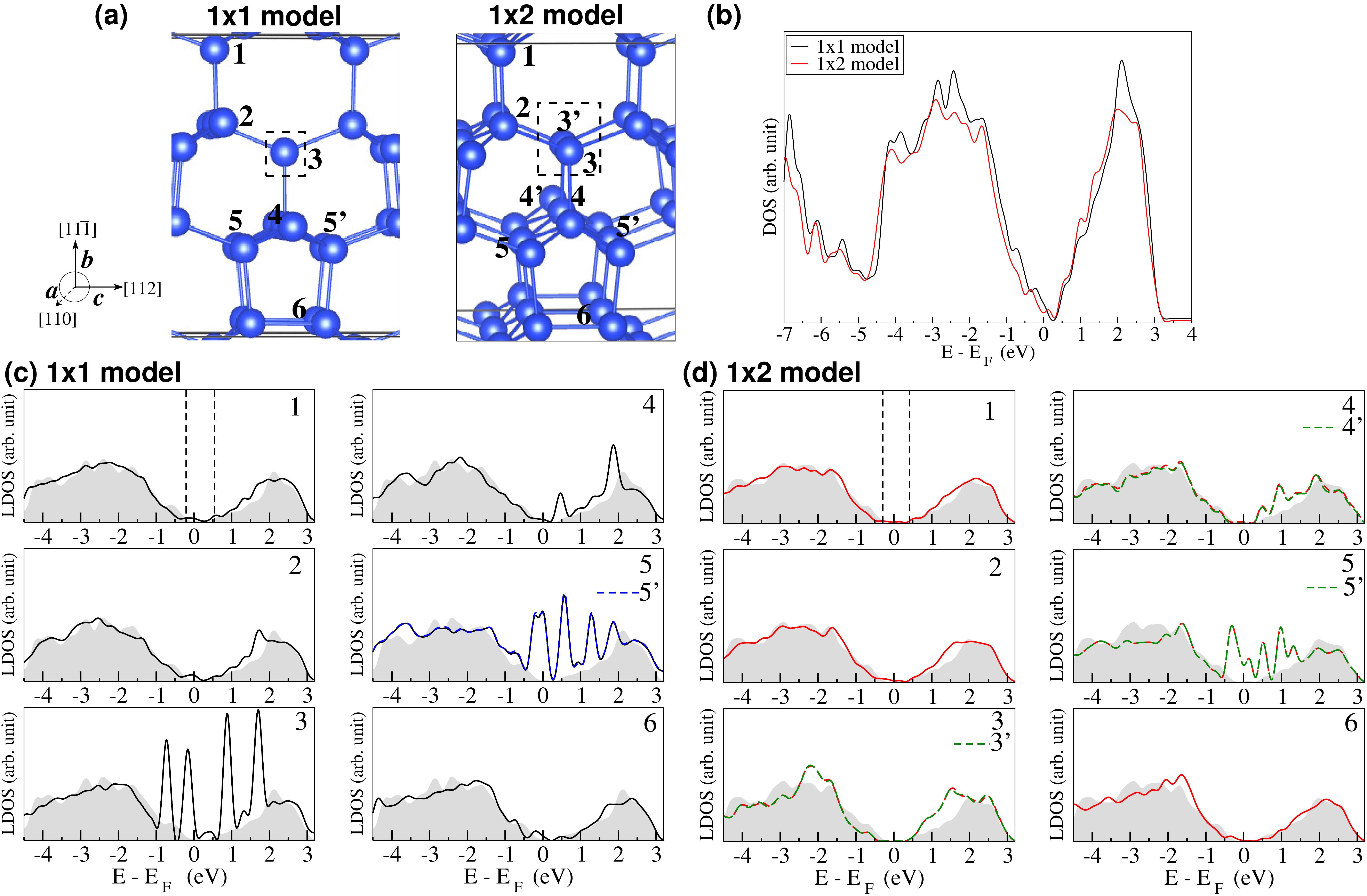}
\caption{GB structure of symmetric $\Sigma$3\{112\} with (a)(1$\times$1) and (1$\times$2) model.
Black square indicate the atomic site in GBs which have dangling bond in the (1$\times$1) GB model and the corresponding dimer in (1$\times$2) model.(b) Density of states (DOS) for (1$\times$1) and (1$\times$2) GB model. Projected local density of states (LDOS) of different sites as labeled within the inset according to (a) for (c)(1$\times$1) and (d)(1$\times$2) GB model. Valence and conduction band edges are marked with dotted lines considering a Si atom from bulk region (gray shade) of the two GB models.}
\label{GB112}
\end{figure*}
%============================================================================

\subsection{The $\Sigma$3\{112\} Si GB (1$\times$1) and (1$\times$2) model structures}
\label{sigb}

To describe the $\Sigma$3\{112\} Si-GB we used a bi-crystal supercell with two GBs and two Si grains, misoriented one with respect to the other by an angle of $70.53^{\circ}$ which form an interface along the crystallographic plane \{112\}. To describe the grain, we used an orthorhombic supercell ($a$ $\ne$ $b$ $\ne$ $c$ and $\alpha$ = $\beta$ = $\gamma$= $90^{\circ}$) and the models (1$\times$1) and (1$\times$2). For the (1$\times$1) model the  lattice parameters are $a$=3.85~\AA, $b$=9.47~\AA~and $c$=80.70~\AA and the cell contains 144 atoms. In the case of the (1$\times$2) model the optimized lattice parameters are $a$=7.65~\AA, $b$=9.48~\AA~and $c$=38.36~\AA ~ and the cell contains 136 atoms. \cite{GBStudio2006,CHENG201892} The GB regions are shown in Fig.\ref{GB112}(a).  

Depending on the initial supercell geometry different structural reconstruction occurs, which lead to distinct local structures in the GBs. For (1$\times$1) model [Fig.\ref{GB112}(a)], sites 1, 2, 5, 5' and 6 all are four coordinated, while Si at site 3 is under coordinated and site 4 is over coordinated. Moreover, the distance between the site 5 and it's mirror symmetric site 5' is 2.84 \AA{} which implies a weak interaction [section \ref{sec:impurity}]. In the case of (1$\times$2) model, the presence of a second layer along the [1$\bar{1}$0] direction saturates the dangling bond of site 3 via dimerization (3 - 3') upon reconstruction. The Si-Si dimer bond length is 2.47 \AA{}, slightly larger than the Si bulk bond length which is 2.35 \AA{}. The other sites 1, 2, 4, 4', 5, 5' and 6 in the (1$\times$2) model GB have the same coordination as the GB (1$\times$1) model, i.e. 4 and 4' both are over-coordinated and distance between 5, 5' is  2.87 \AA{}.

This optimized ground state geometry for the (1$\times$2) model cannot be directly obtained by a standard DFT structural optimization. In fact, the system tend to relax in a local minimum energy configuration in which the Si dimer along [1$\bar{1}$0] is not present. To induce relaxation through the real minimum energy structure special treatments are necessary as, for example, by initially forcing a small Si-Si distance between 3-3' and then reoptimizing the structure. We therefore sought to define a general and reliable protocol that would provide the correct structural configuration without having to resort to manual modifications or trial and error approaches.
Starting from the supercell geometry relaxed through the DFT approach, we performed a molecular dynamics (MD) simulation with VASP. 
The system have been heated up to 300K for 10 ps, and then quenched to 0K using standard MD simulation along with the Nose-Hoover thermostat. Finally the obtained structure has been again optimized using DFT relaxation procedure.
The considered super cell is sufficiently large to use a single k-point for MD calculations.
Through this protocol it was possible to directly obtain the real minimum energy geometries for both model of $\Sigma$3\{112\} Si-GB in accordance with the prediction of \cite{Sawada2006, JAP_H2009, ZHAO2019_AM}.

\section{Results and discussion}\label{sec:result}

\subsection{Energetics and electronic properties: role of local structures}

%\subsection{$\Sigma$3\{112\} Si GB (1$\times$1) and (1$\times$2) models}
  
The formation energy of $\Sigma$3\{112\} Si GB is
\begin{equation}
E^{\text{f}}_{\text{GB}}  = \frac{E_{\text{GB}} - n_{\text{GB}}E_{\text{B}}} {2 A }, 
\label{EfGB} 
\end{equation}
where $E_{\text{GB}}$ is the energy of the GB, $n_{\text{GB}}$ is the number of Si atoms, here 144 and 136 depending on (1$\times$1) and (1$\times$2) model respectively in the GB supercell, $E_{\text{B}}$ is the energy per Si atom taken from the Si cubic crystal structure, $A$ is the GB cross-section area and the scaling factor 1/2 take care of bi-crystal supercell.\cite{MAJI2021116477} 

%-----------------------------------------------------------
We find that the (1$\times$2) model has smaller formation energy ($E^{\text{f}}_{\text{GB}}=0.64$ J/m$^{2}$) than the
corresponding (1$\times$1) ($E^{\text{f}}_{\text{GB}}=1.05$ J/m$^{2}$) model. This is a consequence of the fact that in (1$\times$2) model no dangling bonds are present as a result of reconstruction along [1$\bar{1}$0], thus energy reduces, while in the later case, because of the periodicity of the supercell, three coordinated Si sites are left as marked by black square in Fig.\ref{GB112}(a).

Symmetric $\Sigma$3\{112\} GB contains five and seven-fold Si rings, which constitutes the dislocation core. However, the differences in Si coordinations between the two models induce different local contributions that can be observed in the local density of states (LDOS) in Fig.[\ref{GB112}(c) and (d)], respectively.
In Fig.[\ref{GB112}(c)], for (1$\times$1) model of Fig.\ref{GB112}(a), sites 1, 2 and 6 are all four coordinated and no changes in LDOS appear. Mirror symmetric sites 5 and 5' are still four coordinated but in this case the existence of a weak interaction is reflected by the new peaks in the LDOS gap. Si at site 3 and 4 with dangling bond in the former and presence of extra bond in the later are instead responsible for the extra peaks in the band edges region.  

In case of (1$\times$2) model, the gap states due to site 3, present in the (1$\times$1) model, vanish because of Si dimer formation between 3-3' (see Fig.\ref{GB112}(d)).
All other sites (1, 2, 4, 4' 5, 5' and 6) in the GB 
follows the coordination alike (1$\times$1) model as evident from corresponding LDOS plots.    
Thus reconstruction of bonds locally modifies their contributions to the states, although the total density of states is apparently similar for both the GB models [Fig.\ref{GB112}(b)]. 

\subsection{Vacancy defect}\label{sec:vacancy}
The formation energy of vacancy is calculated as, 
\begin{equation}
E^{\text{f}}_{\text{VGB}}  = E_{\text{VGB}} - \frac{n_{\text{GB}} - 1}{n_{\text{GB}}} E_{\text{GB}} , 
\label{EfVGB} 
\end{equation}
where $E_{\text{VGB}}$ is the energy of the optimized GB including vacancy, $n_{\text{GB}}$ is the number of Si atoms in the supercell, and $E_{\text{GB}}$ is the energy of GB.
%---------------------------------------------------------------
\begin{table}[h!]
\begin{center}
\begin{tabular}{ |c|c|c| } 
\hline
GB model      & vacancy configurations & $E^{\text{f}}_{\text{VGB}}$ (eV) \\
\hline
(1$\times$1)  &  V1 (site 4)      &   0.41 \\ \cline{2-3}
              &  V2( site 5)      &   0.83 \\
\hline
(1$\times$2)  &  V1 (site 4')      &   0.49 \\ \cline{2-3}
              &  V2 (site 5')      &   0.68 \\
\hline              
\end{tabular}
\end{center}
\caption{Vacancy formation energy for V1 and V2 in both the models (1$\times$1) and (1$\times$2) calculated through Eq.\ref{EfVGB}. Si site for creating vacancy configurations V1 and V2 are respectively site 4 and site 5 as marked in Fig.\ref{EfGB}(a).}
\label{energy_val}
\end{table}
%----------------------------------------------------------------------
The values of formation energy in Table \ref{energy_val} suggest that
vacancy V1, due to the removal of a Si atom from the head of pentagon (site 4/4'), is more favorable than V2 (Si removed from site 5/5'), since Si vacancy at a over coordinated site prefers the vacancy segregation which reduce the GB energy. However, here V1 and V2 both within (1$\times$1) and (1$\times$2)
model have been analysed rigorously in order to understand local reconstruction and corresponding effects on electronic properties.

%================================================================================
\begin{figure}[t]
\centering
\includegraphics[scale=0.09]{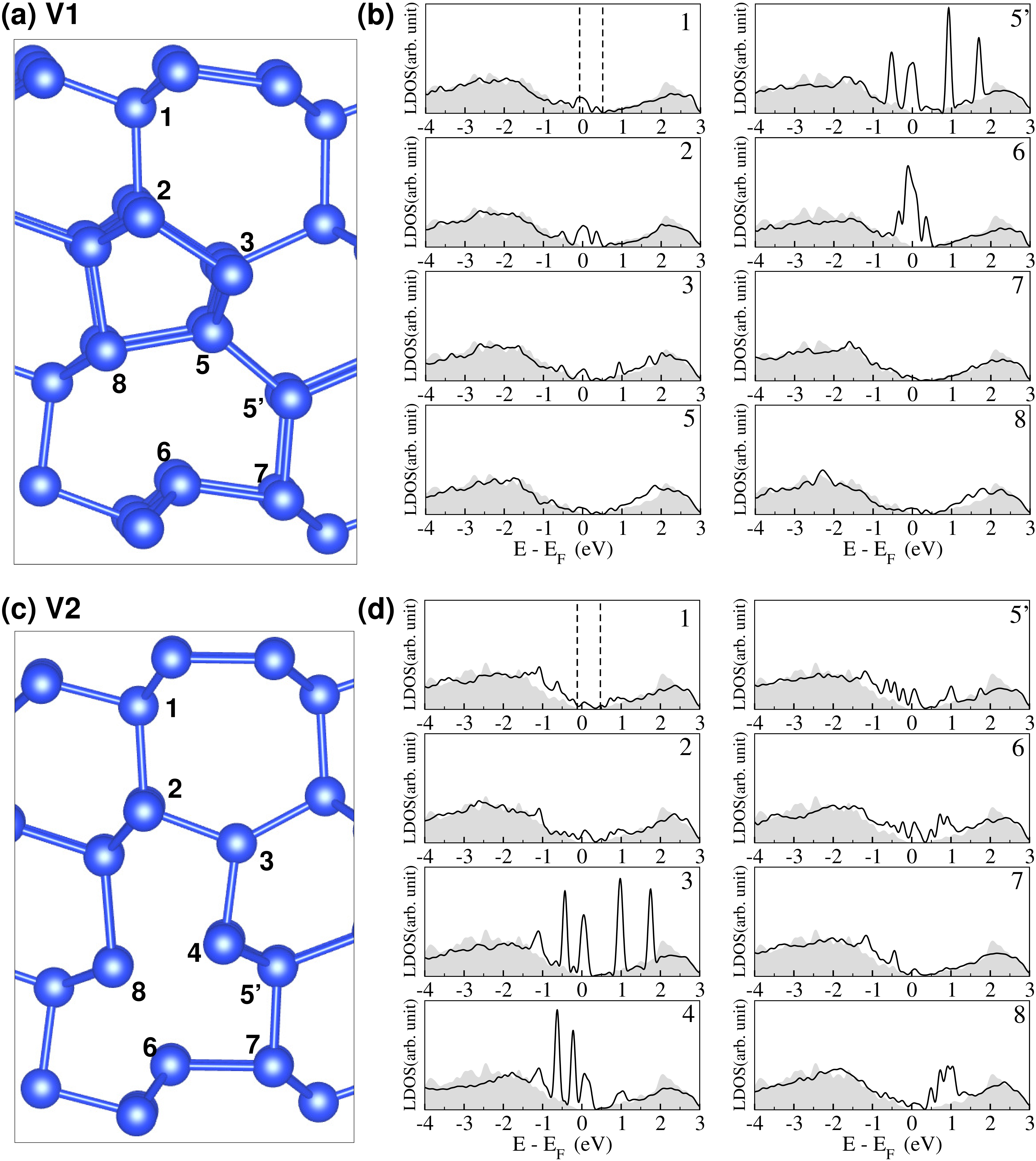}
\caption{Relaxed $\Sigma$3\{112\} Si-GB (1$\times$1) in presence of a vacancy in (a) site V1 or (c) site V2. (b) LDOS plots for sites as marked in (a), (d) LDOS plots for sites as marked in (c). Dotted lines indicate valence and conduction band edges for a Si atom from bulk region (gray shade) of GB models.}
\label{vgb_1x1}
\end{figure}
%=================================================================================
\subparagraph{\bf{1$\times$1 model}}
For vacancy configurations V1 and V2 the reconstructed GB structures are shown in Fig.\ref{vgb_1x1}(a,c).
In the relaxed geometry of V1, the site 3 at GB was three coordinated, and now it undergoes reconstruction and lead to a different dislocation core. Thus except site 5' and 6, that are now three-fold coordinated, Si atoms in all other sites, as marked in Fig.\ref{vgb_1x1}(a) are now four-fold coordinated.
The presence of dangling bonds due to site 5' and 6 provide new states in the gap as evident from LDOS [Fig.\ref{vgb_1x1}(b)].
However, for V2, no reconstruction occur for site 3 and the new optimized core leaves more sites (4, 6 and 8) with dangling bond. Therefore the presence of states in the gap increases, with a major contribution from sites 3 and 4. For both V1 and V2, some small peaks in the LDOS gap are visible also for other sites, due to the lattice distortion induced by the reconstruction needed to accommodate the vacancy

\begin{figure}[ht]
\centering
\includegraphics[scale=0.11]{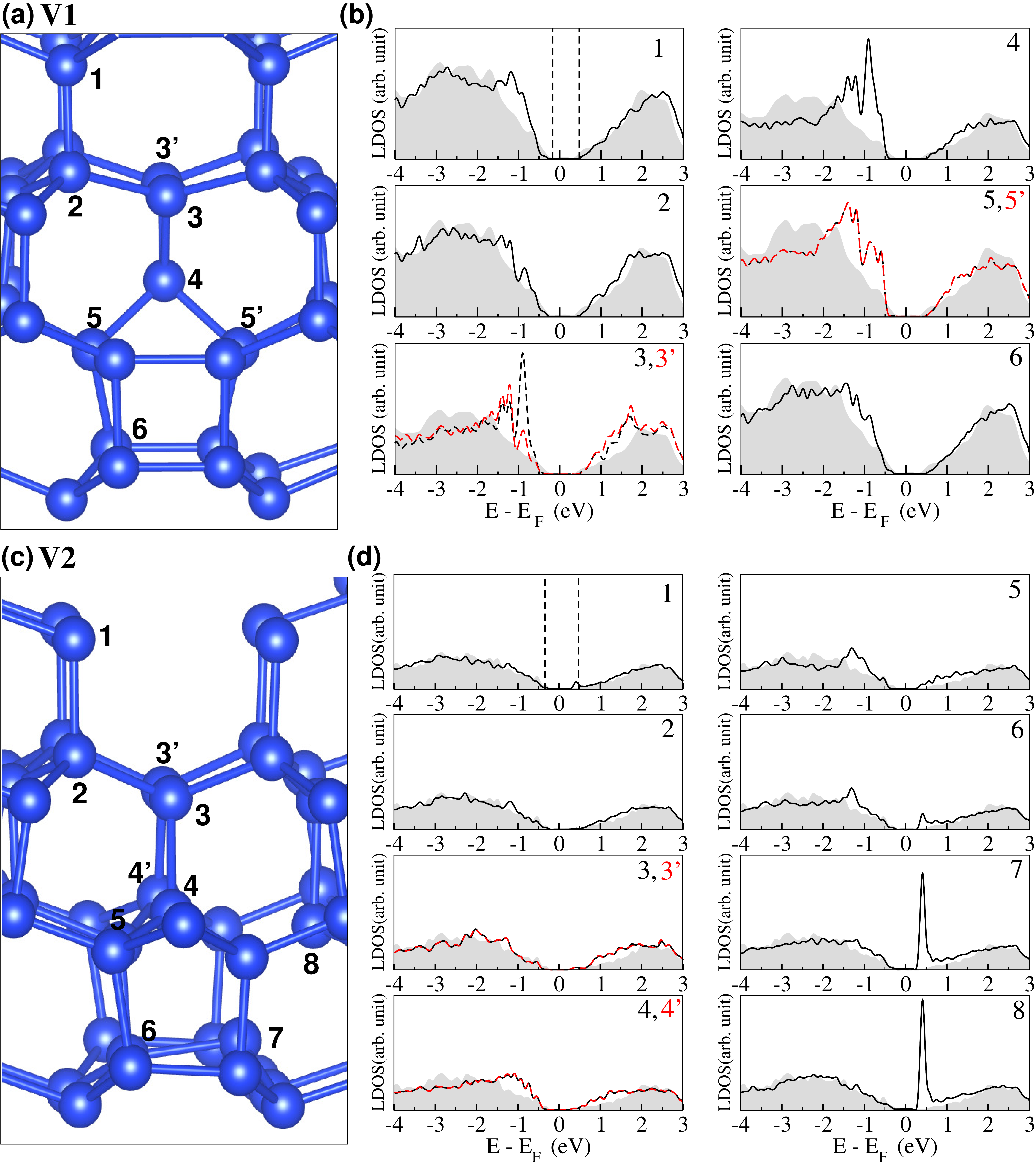}
\caption{Relaxed $\Sigma$3\{112\} Si-GB (1$\times$2) in presence of a vacancy in (a) site V1 or (c) site V2. (b) LDOS plots for sites as marked in (a), (d) LDOS plots for sites as marked in (c). Dotted lines indicate valence and conduction band edges for a Si atom from bulk region (gray shade) of GB models.}
\label{vgb_1x2}
\end{figure}

\subparagraph{\bf{1$\times$2 model}}
In bulk Si, the presence of a vacancy represents a point defect that induce dangling bonds in the system. $\Sigma$3\{112\} Si-GB (1$\times$2) model, as bulk Si, does not present any dangling bonds, nevertheless a neutral vacancy is perfectly segregated.
Actually, structural optimization of
vacancy defect V1 in the GB resulted again in a structure perfectly bonded, without dangling bonds, that could be seen
as another possible configuration of the same grain boundary.\cite{JAP_H2009} In LDOS [Fig.~\ref{vgb_1x2}(a, b)], absence of any defect states in the gap confirms the disappearence of coordination defects at the GB.\\
Vacancy V2 leads, instead, to coordination defects for sites 7 and 8, however in optimized structure these two sites tend to move closer to each other. This pair partially restore the four-fold coordination, thus shifting local levels towards conduction band edges [Fig.~\ref{vgb_1x2}(c, d)].
%========================================================================================
%==========================================================================================

\subsection{Segregation: H, C, N and O impurities }\label{sec:impurity}
The formation of Si vacancies at GB dramatically alters the atomic structure and thus the electronic properties and passivation behaviors of GBs in Si. Therefore, segregation of different impurities have also been affected. Here we consider single H, C, N and O atoms as segregated elements, in both (1$\times$1) and (1$\times$2) model. The segregation energy of the impurity atom at different interstitial positions at the GB in presence of a vacancy\cite{MAJI2021116477} is computed as:
\begin{equation}
\Delta^{\text{XVGB}}_{\text{XGB}} = E^{\text{XVGB}} - E^{\text{XGB}} 
\label{deltaXVGB} 
\end{equation}
where X refers to different impurity elements (H, C, N and O here). $E^{\text{XVGB}}$ and $E^{\text{XGB}}$ are the impurity energies respectively in the GB with a vacancy and in the pristine GB which are calculated as
\begin{equation}
E^{\text{XVGB}} = E_{\text{VGB+X}}- E_{\text{VGB}}  - \mu_{\text{X}}, 
\label{EnXVGB} 
\end{equation}
\begin{equation}
E^{\text{XGB}} = E_{\text{GB+X}} - E_{\text{GB}}  -  \mu_{\text{X}},
\label{EnXGB} 
\end{equation}
where $E_{\text{VGB+X}}$ is the energy of the GB with a vacancy along with impurity atom (X: H/C/N/O), $E_{\text{VGB}}$ is the energy of the GB with a vacancy, $\mu_{\text{X}}$ is the chemical potential of the respective impurity atom, and $E_{\text{GB+X}}$ is the total energy of the GB containing the impurity atom, $E_{\text{GB}}$ is the total energy of the Si GB.

%%-------------------=====================1x1====================----
%=========================================================================================
\begin{figure*}[ht]
\centering
\includegraphics[scale=0.13]{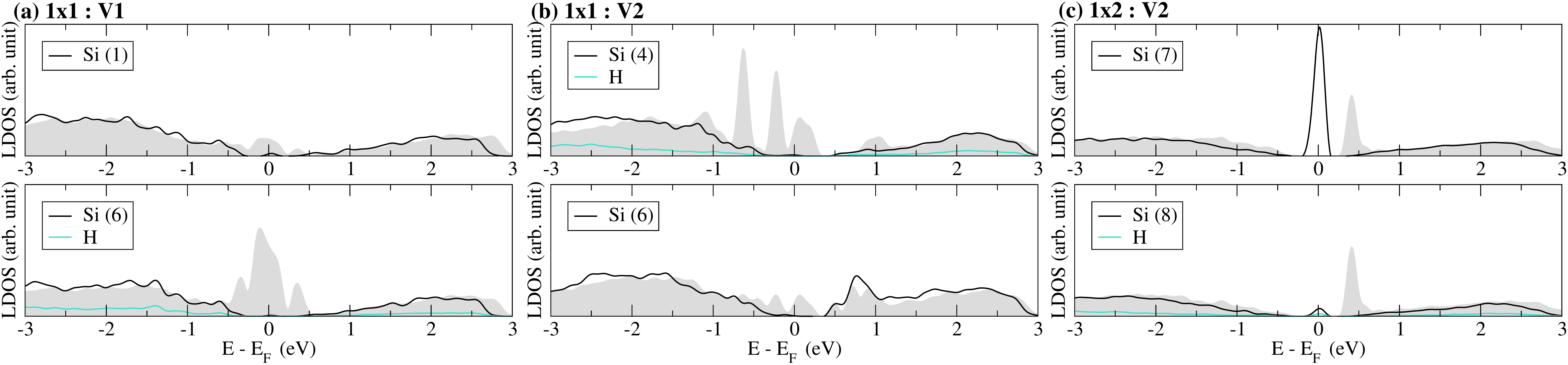}
\caption{LDOS for Si and H sites for (a)(1$\times$1) model with vacancy V1, (b)(1$\times$1) model with vacancy V2, and (c)(1$\times$2) model with vacancy V2. In the inset site numbers are in accordance with Fig.\ref{vgb_1x1}(a,c) and Fig.\ref{vgb_1x2}(c). LDOS of H is plotted along with that of Si atoms that it saturates. Same sites before introducing the H impurity are plotted with grey shade.}
\label{ldos_H}
\end{figure*}
%====================================================================

Concerning the electronic properties, we have chosen to discuss here the structures with the lowest segregation energy configurations. We have reported the LDOS of the sites as marked in presence of vacancy in GB for both the models [Fig.\ref{vgb_1x1}(a,c) and Fig.\ref{vgb_1x2}(c)]. The total DOS can be found in the Supplementary Material [SM]. 

A single H atom can saturate only one of the 3-fold coordinated sites. 
However in both the models (1$\times$1) and (1$\times$2), the presence of 
a Si vacancy corresponds to more than one unsaturated Si, hence, single H atom segregation cannot fully passivate the GB. %as evident from presence of defect states in LDOS.
Si at site 6 in (1$\times$1) model with vacancy V1 [Fig.\ref{ldos_H}(a)], Si at site 4 in (1$\times$1) model with vacancy V2 [Fig.\ref{ldos_H}(b)] and Si at site 8 in (1$\times$2) model with vacancy V2 [Fig.\ref{ldos_H}(c)] are passivated by the H atom resulting in the disappearance of gap states in the two former cases and with the lowering and shift inside the gap in the latter. In this last case the inclusion of H increase the distance between Si(7) and Si(8) discarding the partial coordination restoration leading to more localized states. The H passivation also reduces local distorsion of neighbouring sites, thus removing low lying states for Si(1) in Fig.\ref{ldos_H} (a) and Si(6) in Fig.\ref{ldos_H} (b).

\begin{figure*}[t]
\centering
\includegraphics[scale=0.18]{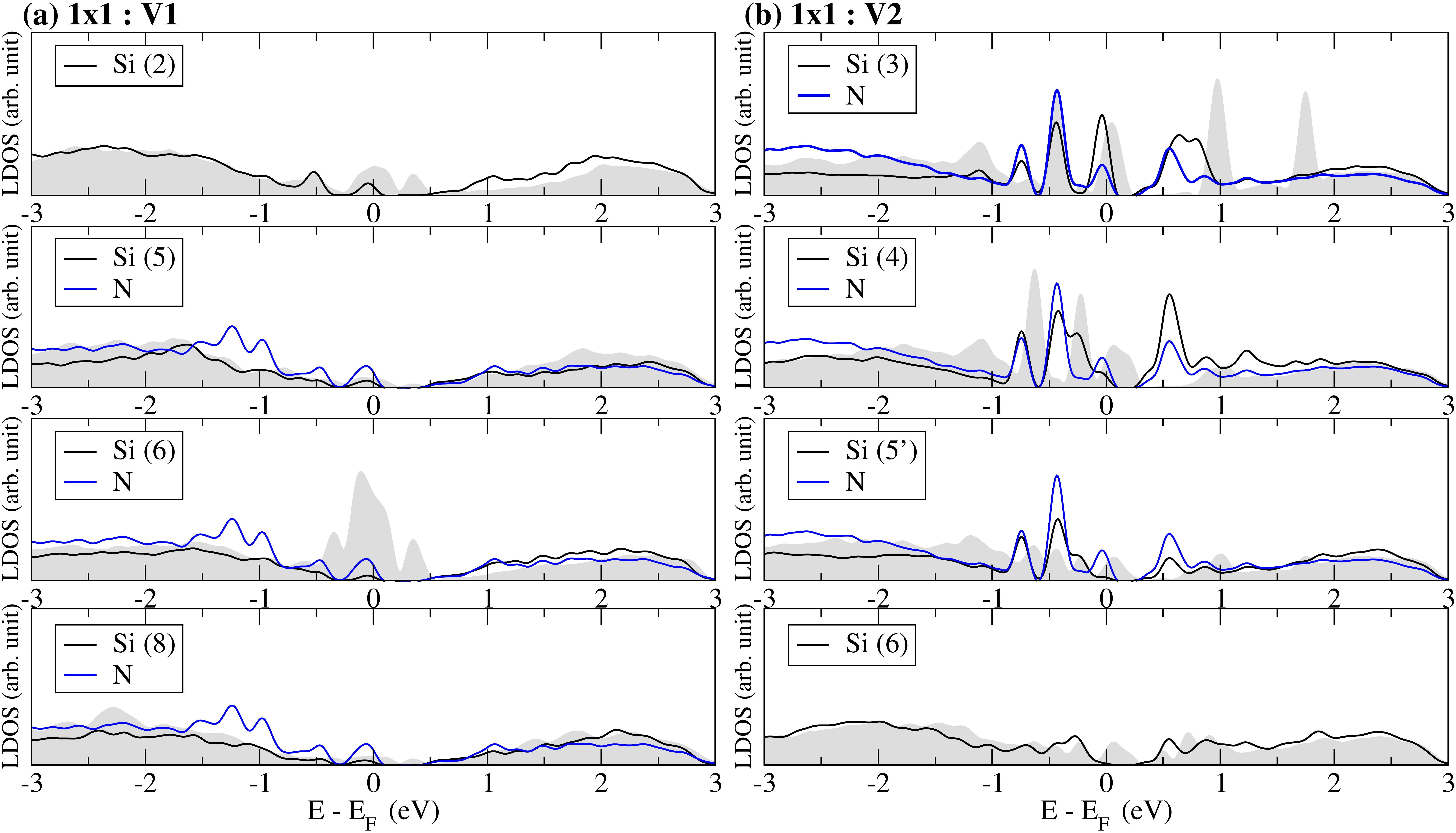}
\caption{LDOS for Si and N sites for (1$\times$1) model with vacancy (a)V1 and (b)V2. In the inset site numbers are in accordance with Fig.\ref{vgb_1x1}(a,c). LDOS of N is plotted along with that of Si atoms that it saturates. Same sites before introducing the N impurity are plotted with grey shade.}
\label{ldos_N}
\end{figure*}

The segregation characteristics of the C, N and O atoms differ greatly when they are placed as interstitial impurities on the two GB models. In this process even the lowest segregation energy configurations confront with under or over coordinated situations. 
In Fig.\ref{ldos_N}(a,b) we have considered two configurations, where N fulfil it's coordination. 
Although in both cases the N atom is three-fold coordinated and Si atoms bonded with N are four-fold coordinated, the situation is substantially different. Actually, while the LDOS of 1x1 model with vacancy V1 shows the effective role played by the N atom in cleaning the gap from states due to dangling bonds (see for example LDOS of atoms 5, 6 and 8 of Fig.\ref{ldos_N}(a)), the LDOS of 1x1 model with vacancy V2 (Fig.\ref{ldos_N}(b)) present more states in the gap for atoms 3, 4 and 5'. These states are due to structural distortions of the lattice: actually, while the system with V1 vacancy, in the presence of N, recovers a certain symmetry, the model with V2 maintains a high distortion at the GB. 
On the contrary, for Si at site 6 [Fig.\ref{ldos_N} (b)], that through reconstruction is now 4-fold coordinated and present a quite symmetric configuration around, the peak in the gap vanishes.

\begin{figure*}[t]
\centering
\includegraphics[scale=0.18]{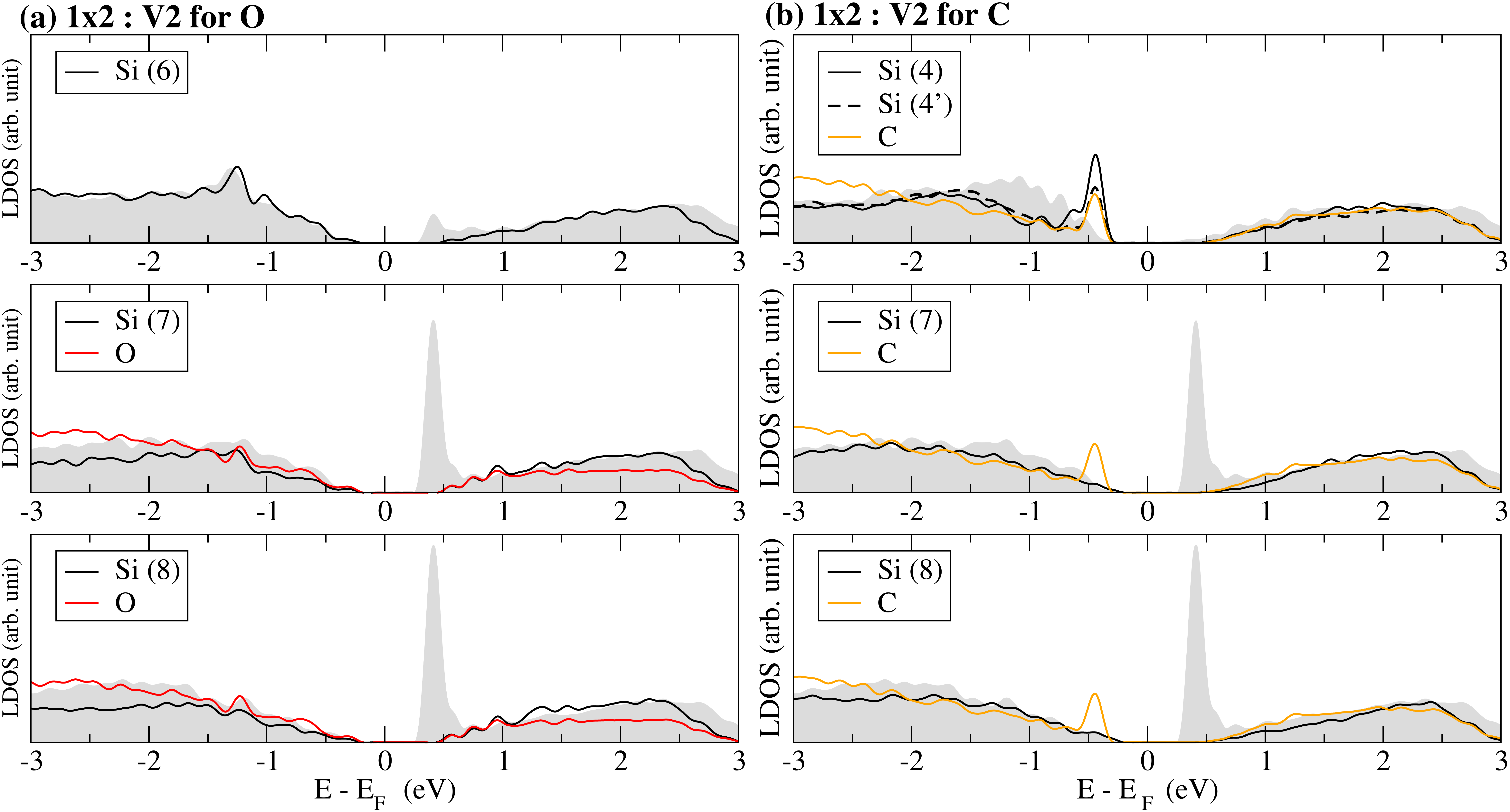}
\caption{LDOS for (1$\times$2) model with vacancy V2 and  (a) O impurity, (b) C impurity. In the inset site numbers are indicated according to Fig.\ref{vgb_1x1}(c) and Fig.\ref{vgb_1x2}(c). LDOS of O/C is plotted along with that of Si atoms that the impurity saturates. Same sites before introducing the O/C impurity are plotted with grey shade.}
\label{ldos_CO}
\end{figure*}

%%------=================1x2==================------
In Fig.\ref{ldos_CO}, we show the LDOS of Si and interstitial C and O atom for (1$\times$2) model in the presence of vacancy V2.
O atom segregates at bond-centered configuration between the two 3-fold coordinated Si atoms (site 7 and 8 in [Fig.\ref{vgb_1x2}(c)]) restoring local Si-O-Si structure as happen in Si bulk and Si nanostructures\cite{MAJI2021116477,JNNDegoli2008,DEGOLI2009203} as confirmed by the vanishing of gap states [Fig.\ref{ldos_CO}(a)].
Moreover, the inclusion of the O atom reduces the lattice distortion induced by the vacancy as is demonstrated by the LDOS of Si(6)
that, without gap states, is now quite similar to the initial one [see Fig.\ref{GB112}(d)].

For interstitial C segregation, C atom as well as the Si atoms in the GB
restore their 4-fold coordination. From LDOS plots [Fig.\ref{ldos_CO}(b)] of Si sites as marked in Fig.\ref{vgb_1x2}(c), and C atom (included in the plot of the four Si atoms with which it is bonded), it is evident that while the peak at the conduction band edge (atom 7 and 8 grey shade),  vanishes due to passivation of dangling bonds, a new peak appears at the valence band edge probably due to the lattice distortion induced by C atom.
Present calculations actually show that interstitial C atoms can lead to a severe lattice distorsion of surrounding Si crystal, as a result of the formation of longer C-Si covalent bonds (1.956\AA) in comparison to most stable bond in bulk SiC (length 1.887\AA) \cite{Dong2004}. 
%====================================================================
%====================================================================
%====================================================================
%==============================================================================

\begin{figure*}[t]
\centering
\includegraphics[scale=0.115]{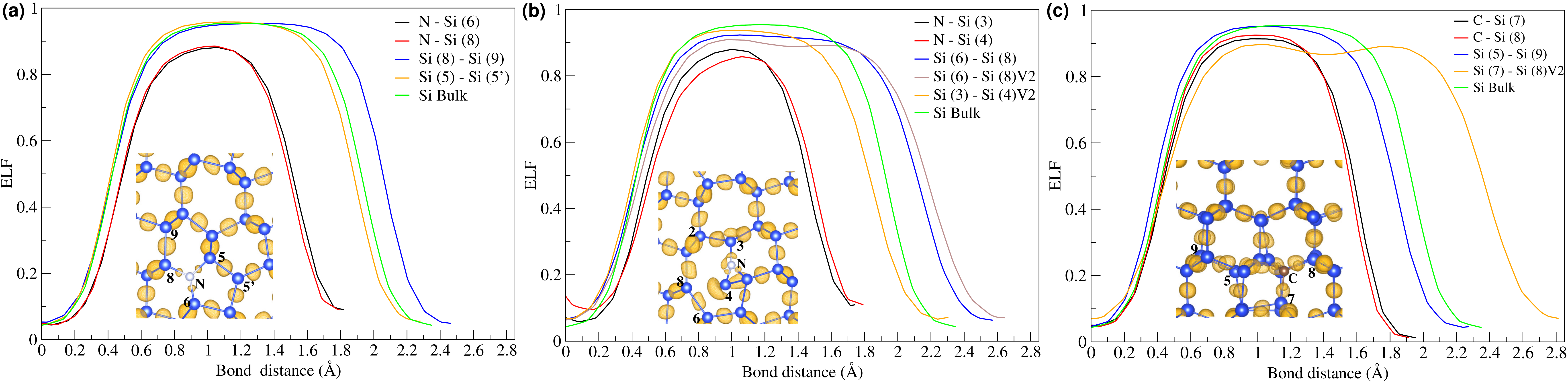}
\caption{ELF iso-surface plot (in yellow) (inset, isovalue=0.85) for (a) N in 1$\times$1 model with V1 (b) N in 1$\times$1 model with V2 and (c) C in 1$\times$2 model with V2. Si atoms are blue, the N atom is white and C atom is brown. Coloured lines represents ELF profiles along particular bonds, as described in the legend, for configurations (a), (b) and (c).  The label V2 in the legend refers to bonds, taken for comparison, of the structures with the vacancy only and not the impurity. As reference, ELF line profile in bulk Si is always reported in green.}
\label{elf}
\end{figure*}
%==============================================================================
\begin{figure}[b]
\centering
\includegraphics[scale=0.32]{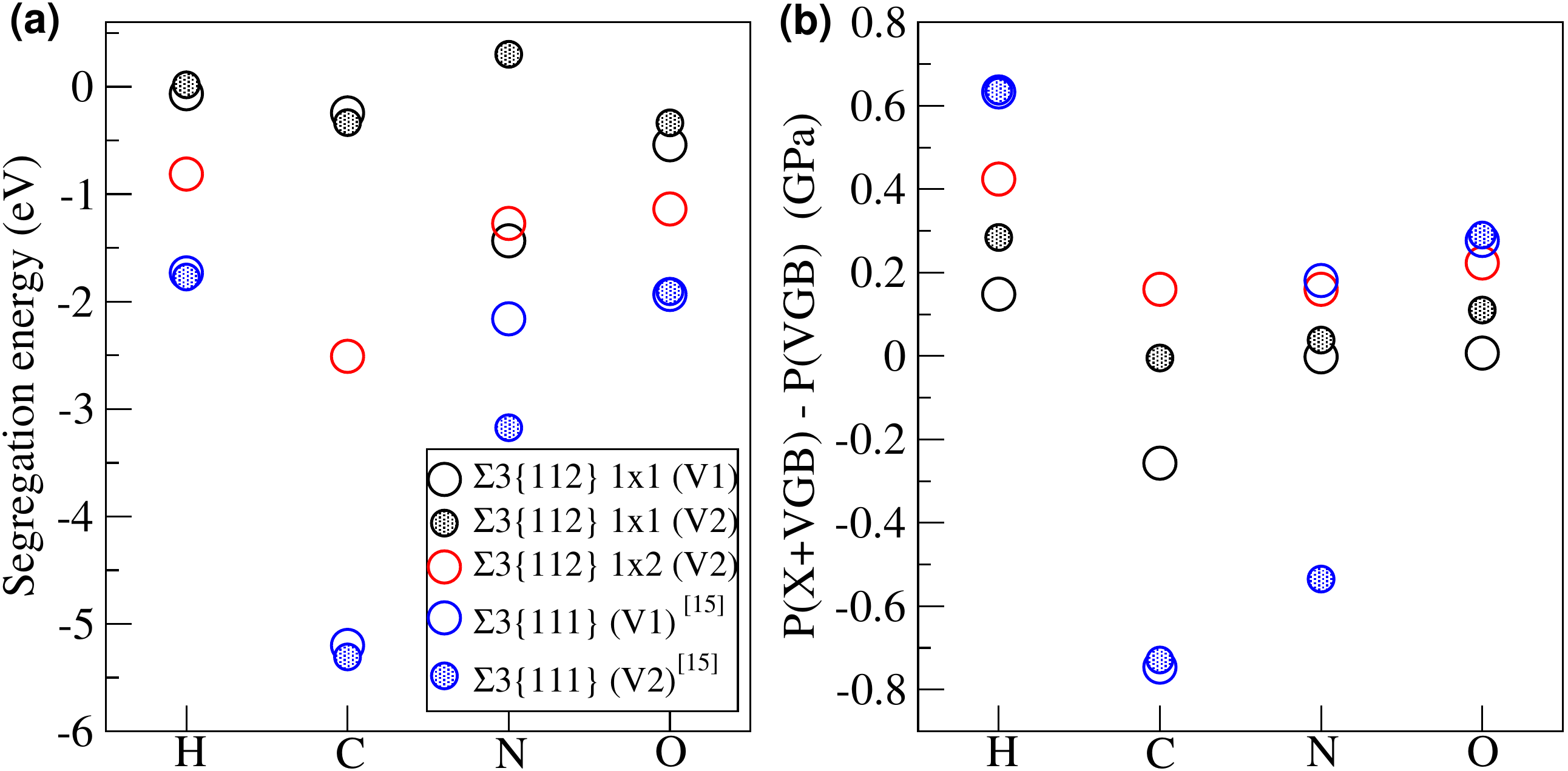}
\caption{Comparison of (a) segregation energy and (b) system pressure (P(X+VGB) - P(VGB)) in presence of different interstitial impurities X (X: H, C, N, O) and vacancy defect (VGB) on two different GB model of $\Sigma 3\{112\}$, compared with the two possible vacancy defect structures on most stable $\Sigma 3\{111\}$ GB\cite{MAJI2021116477} as mentioned in the inset. Values of the segregation energies and pressures are also reported in Table.1 in the Supplementary Material.}
\label{vgb_compare}
\end{figure}
%================================================================================
For in-depth comprehension of the complex configurations obtained, we resort to the analysis of bonding structure of segregated impurities obtained from electron localization function (ELF)\cite{elf1994}. In Figure Fig.\ref{elf}, ELF profile is plotted along different bonds as marked in the inset. ELF is a relative measurement of the electron localization and it takes values between 0 and 1. We note
that in the PAW method only valence electrons are taken into
account in the calculation so ELF profile starts from zero in the core of the atoms. When the amplitude of ELF is $\approx$1, then electrons are localized in the bonding
regions, as observed from typical shape of the covalent bond in bulk Si [green line in Fig.\ref{elf}(a-c)]. ELF profiles in Fig.\ref{elf}(a-c) show the strained nature of Si-Si bonds, either elongated or compressed, due to segregated N/C atom. The reduced amplitude of ELF for Si-N and Si-C bonds [red and black lines in Fig.\ref{elf}(a, b, c)] indicates the shorter heteropolar covalent bond: ELF amplitude for C-Si bond is closer to Si-Si bond than Si-N bond, indicating the variety of heteropolar covalent bonds. 
The relative difference between ELF profile for different Si-N bonds in Fig.\ref{elf}(b) with respect to Fig.\ref{elf}(a) indicates the more distorted structure with V2 rather than V1.
Moreover ELF profile along the three folded Si atoms [Si(6)-Si(8) V2 in Fig.\ref{elf}(b) and Si(7)-Si(8) V2 in Fig.\ref{elf}(c)] shows the  delocalized nature of these bonds (visible in the lowering of the central part of ELF profile) due to a weak interaction as discussed previously. \\
Moreover, these data are confirmed by the ELF iso-surface plot in the sketch that is regularly (i.e. almost spherical) distributed on localized Si-Si bonds while it is more and more distorted in presence of dangling/delocalized bonds (see for example Fig.\ref{elf}(b) around N and around Si(4)).

\subsection{$\Sigma$3\{112\} vs  $\Sigma$3\{111\} Si grain boundaries}\label{sec:discussion} %GBs comparison}	
%%ADD OBSERVATION FOR  $\Sigma 3\{112\}$ and compare 
The segregation mechanisms depend on local structural distortions obtained by the GBs itself or otherwise generated by
strain and/or vacancies. The most symmetric and stable $\Sigma 3\{111\}$
Si-GB is well known to be less effective in pristine state, instead it favores segregation in presence of vacancy defects\cite{MAJI2021116477,JCP_Rmaji2021}. However, less symmetric boundaries such as $\Sigma 3\{112\}$, $\Sigma 9\{221\}$ and $\Sigma 27\{552\}$ demonstrated to be efficient gettering centers for atomic impurities in pristine structure itself.\cite{ZHAO2018,ZHAO2019_AM,ZHAO2017599,PhysRevLett_121_015702}

$\Sigma 3\{112\}$ has been discussed for both symmetric and asymmetric GB structures, in term of their gettering ability \cite{ZHAO2018}. However for symmetric $\Sigma 3\{112\}$ Si-GB, considerations on the supercell lead to two different models and in presence of vacancy the optimized geometry correspond to different segregation energetics.\\
In Fig. \ref{vgb_compare} we show the segregation energies (a) and the pressure variation for the $\Sigma 3\{111\}$\cite{MAJI2021116477,JCP_Rmaji2021} and $\Sigma 3\{112\}$ Si-GBs in the presence of a vacancy as a function of different interstitial impurities (H, C, N, and O) choosing the minimum energy configuration in between many possible equilibrium structures.

The behaviour is very different both for the particular GB and for the different interstitial impurities. 
In general we can see that, in presence of a vacancy, all the considered impurities segregate much better in $\Sigma 3\{111\}$  than in $\Sigma 3\{112\}$ GB and (1$\times$1) model is, among all, the least effective in segregation (i.e. impurities are not able to segregate).
Both in $\Sigma 3\{112\}$ (1$\times$2) model and $\Sigma 3\{111\}$ GB, C segregation is favored followed by N, O and H eventually. 
Moreover, the segregation energy is only slightly affected by the vacancy location, except for the case of N atom where a strong dependence is clear: for example the segregation of N is favoured in (1$\times$1) model with V1 vacancy while it becomes strongly unfavorable with vacancy V2.
The trend in segregation energy is confirmed by changes in the hydrostatic pressure of our structures, which is obtained from VASP for a given cell with a fixed volume.
If we look at the changes in the hydrostatic pressure when a vacancy is created at the GB we find that, while in $\Sigma 3\{111\}$ GB pressure decreases (P(VGB) - P(GB) $\simeq$ -0.48 GPa) indicating, as expected, a lowering of the total stress in the system, in $\Sigma 3\{112\}$ GB the pressure is almost unchanged (P(VGB) - P(GB)$\simeq 0$) probably as a consequence of the strongly asymmetric nature of this GB with respect to the most symmetric $\Sigma 3\{111\}$.

Changes in the hydrostatic pressure when an impurity atom is then added at the GB are shown in Fig. \ref{vgb_compare}(b).
$\Sigma 3\{111\}$ GB experiences a further lowering of pressure, and therefore of structural stress, in the case of C and N (only for the most stable vacancy) impurities. O atom increases the pressure with respect to the GB with a vacancy  but the whole system has anyway lower pressure than the pristine GB. Finally H impurity increases the system pressure above the one of pristine GB indicating a consistent stess at the GB itself.

For the $\Sigma 3\{112\}$ GB, only C impurity reduces the total pressure of the system, while N and O atoms only slightly increase it. H is again the one with higher increase in pressure, actually a single H is not able to stabilize the system both electronically and structurally.

\section{Conclusions}
In the present work, we studied the structural and electronic properties of the 1$\times$1 and the 1$\times$2  $\Sigma 3\{112\}$ Si-GB models via first-principles calculations. We demonstrated that structural optimization plays a major role in obtaining reliable structures and therefore also reliable physical properties of the systems. For this reason, we have defined a general protocol able to avoid the possible local structural minima that DFT can incur, and that would affect the study of the system.
With this approach, we have investigated how vacancies as well as light impurity elements segregate and interact with the GB, and how they change the structural and the electronic properties of the GB.
As a result we have found a correlation between segregation energies of intrinsic/impurity defects with local geometry of the Si-GB models. \\
We have found that the presence of vacancy sites lead to either perfectly bonded Si-GB or to more distorted GB structures depending on the initial supercell model, i.e. (1$\times$2) $\Sigma 3\{112\}$ Si-GB completely segregates the vacancy V1 thus not showing coordination defects and associated states in the gap. The opposite behaviour was found for the (1$\times$1) $\Sigma 3\{112\}$ Si-GB model.\\
In addition, the inclusion of a light impurity of different valency (C, N, H, O) may lead to fully-/un-saturated bonds depending on the nature of the impurity. In the process of segregation, the energy can decreases up to 2.5 eV for $\Sigma 3\{112\}$ Si-GB, and varies for different elements. The defect levels due to the vacancy are shifted towards the conduction band, becoming shallow levels and then, in the case of the most energetically favourable segregated impurity these gap states disappears. At
the same time, the Si atoms along the GB try to relax decreasing the interatomic distances and forming pairs. For the most favourable situations, covalent bonds between these atoms are reconstructed leading to a systems without states in the gap. 
When this reconstruction is not possible due to the structural distortion of the migrating impurity new states inside the gap are observed. \\
Depending on the GB models considered we found that the energetics as well as the ability of impurities in reducing coordination defect and associated gap states, detrimental for the devices, can be different. In the case of (1$\times$2) model, C and O seems to be the more effective impurity elements while in (1$\times$1) model neither C, N or O seem able to fulfil the proper coordination while H results to be the best candidate for full segregation/passivation.

\section{Supplementary Material}
In the Supplementary Material structures and local density of states are 
reported for all sites as marked in the structures. 

\section{Acknowledgements}
We would like to acknowledge the University of Modena and
Reggio Emilia for financial support (FAR dipartimentale 2021) and
Centro Interdipartimentale En\&Tech, as well as the CINECA HPC
facility for the approved ISCRA B Project POLYFACE (Grant No. HP10BVBM14) and ISCRA C Projects SiGB-NMI (Grant No.
HP10CJEGPF) and Interpol (Grant No. HP10CMNPBM).

%\bibliography{sigb_structure1.bbl}
%\bibliographystyle{unsrt}

%%%%%%%%%% END DOCUMENT %%%%%%%%%%
%\begin{thebibliography}{64}

%\end{thebibliography}
%\end{document}

\end{document}